\newcommand{\pt}{$p_{\mathrm t}$\xspace}
\newcommand{\mt}{$m_{\mathrm t}$\xspace}
\newcommand{\roots}{$\sqrt{s}$\xspace}
\newcommand{\rootsn}{$\sqrt{s_{\mathrm{NN}}}$\xspace}
\newcommand{\GeVc}{GeV/$c$\xspace}
\newcommand{\TeV}{TeV\xspace}
\newcommand{\ptrange}[2]{${#1}\ \textrm{GeV}/c\xspace < p_{\mathrm{t}}\xspace < {#2}\ \textrm{GeV}/c$}
\newcommand{\etarange}[2]{${#1} < \eta < {#2}$}
\documentclass[
    ,final            % use final for the camera ready runs
  ]
  {aipproc}

\usepackage{float}
\usepackage{subfig}
\usepackage{xspace}

\layoutstyle{6x9}

\begin{document}

\title{Measurement of the nuclear modification factor of electrons from heavy-flavour hadron decays in 
Pb--Pb collisions at \rootsn = 2.76 \TeV with ALICE at the LHC}


\author{Markus Fasel for the ALICE collaboration}{
  address={Research Division and ExtreMe Matter Institute EMMI, 
  GSI Helmholtzzentrum f\"ur Schwerionenforschung, Planckstr. 1, 64291 Darmstadt, Germany}
}

\begin{abstract}
We present a measurement of the nuclear modification factor of electrons from heavy-flavour hadron decays at midrapidity in Pb--Pb collisions at \rootsn = 2.76 \TeV. Electrons are identified in the \pt range \ptrange{1.5}{6}. A suppression is seen for \pt larger than 3.5 \GeVc in the most central collisions.
\end{abstract}

\maketitle

%%%%%%%%%%%%%%%%%%%%%%%%%%%%%%%%%%%%%%%%%%%%
%% MAINMATTER
%%%%%%%%%%%%%%%%%%%%%%%%%%%%%%%%%%%%%%%%%%%%

\section{Introduction}
\label{sec:intro}
Heavy quarks are produced at the very early stage of a nucleus-nucleus collision. Thus they are a good tool to study partonic energy loss in the hot and dense medium. Due to the dead-cone effect \cite{Dokshitzer2001199} an ordering of the energy loss according to the quark mass is expected leading to a stronger suppression for charm than for beauty \cite{PhysRevLett.94.112301}. In measurements at RHIC \cite{PhysRevLett.98.172301, PhysRevLett.96.032301} a strong suppression of electrons from heavy-flavour is observed for \pt > 2 \GeVc in central collisions. Comparisons to model calculations imply that pure radiative energy loss underpredicts the suppression observed . This opens the possibility for additional energy loss from elastic scattering of the heavy quarks while traversing the medium \cite{Wicks2007493}.  

Heavy-flavour measurements can be performed in the hadronic channels by direct reconstruction of the decay of heavy-flavour hadrons \cite{ortona} and in the semi-leptonic channels. The muonic channel is discussed in \cite{bastid} and the electronic channel is presented here. 

The measurement is performed with ALICE (A Large Ion Collider Experiment \cite{ALICE_PPR1, ALICE_PPR2}) in Pb--Pb collisions at \rootsn = 2.76 \TeV in the pseudorapidity interval $|\eta| < 0.8$.  A minimum bias trigger requiring hits in the Silicon Pixel Detector (SPD), which has a pseudorapidity coverage $|\eta| < 1.4$ for the outer respectively $|\eta| < 2$ for the inner layer, and the VZERO detector with a pseudorapidity coverage \etarange{-3.7}{-1.7} respectively \etarange{2.8}{5.1} is used. The collision centrality is estimated by applying cuts on the amplitude measured in the VZERO. Background collisions are rejected using the timing information of VZERO and the Zero Degree Calorimeter (ZDC). Only events within $\pm$ 10 cm around the nominal interaction point along the beam axis are considered for this analysis. In total $1.7 \times 10^{7}$ events are analyzed. 

\section{Measurement of the nuclear modification factor of electrons from heavy-flavour hadron decays}
Tracks are required to fulfill certain quality cuts to suppress background tracks. In order to suppress electrons from photon conversions in the Inner Tracking System (ITS) and the material at the entrance of the Time Projection Chamber (TPC), tracks are required to have a hit in the first ITS layer, which is the first SPD layer.

\label{sec:elid}
\begin{figure}[t]
\centering
\includegraphics[width=0.7\textwidth]{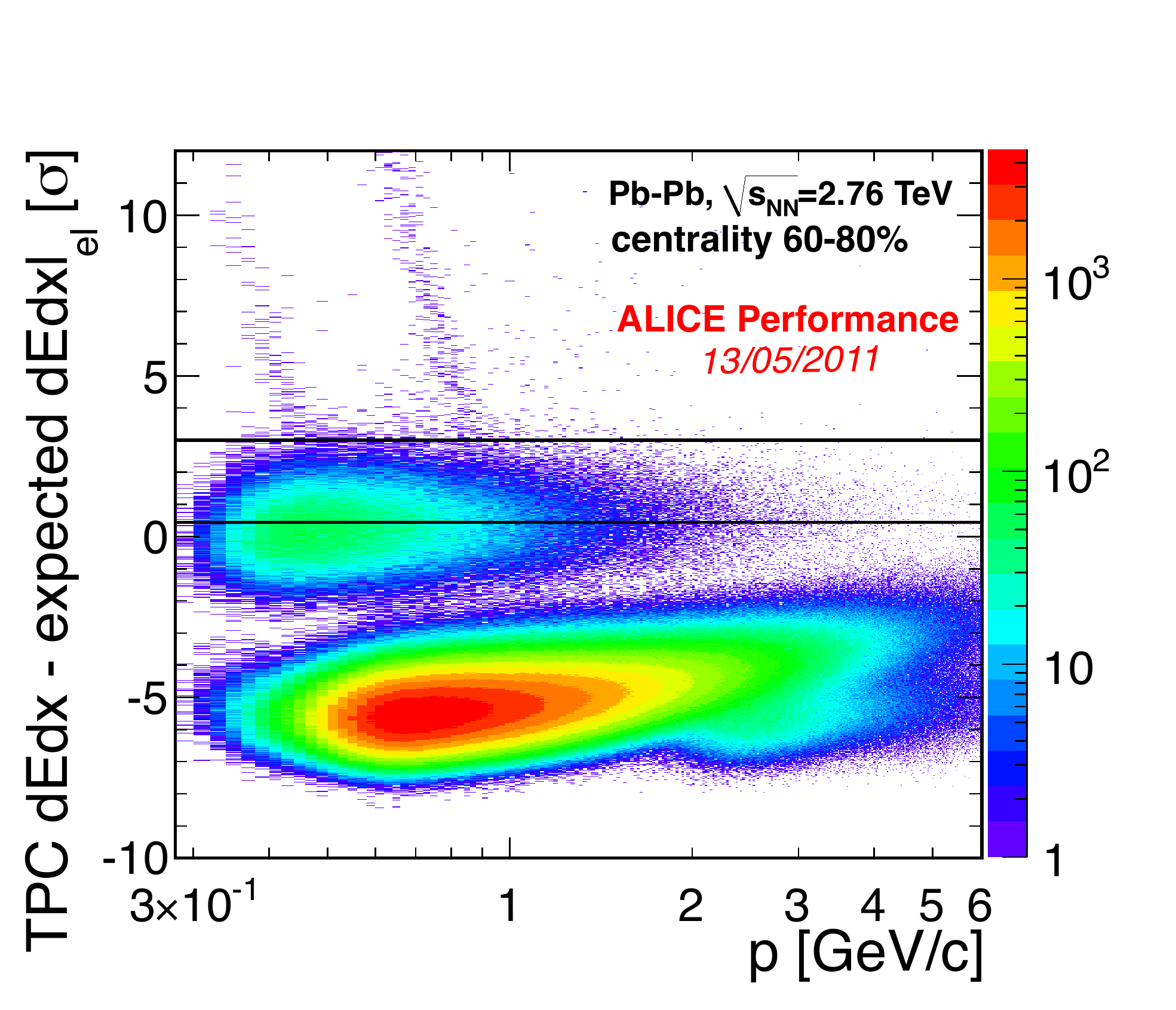}
\caption{Selection of electron tracks in the TPC for the Pb--Pb centrality class 60-80\%: the energy loss expressed relative to the expected energy loss of electrons is shown after TOF as function of momentum in units of standard deviations. Black lines indicate the selection band.}
\label{fig:TPCpid}
\end{figure}
Electrons are identified by applying cuts on the signal in the Time-of-Flight detector (TOF) and the Time Projection Chamber (TPC). In order to reject protons and kaons the measured time-of-flight is required to be within 3$\sigma$ of the expected time-of-flight of electrons with a current time-of-flight resolution of 160 ps. For tracks passing this cut the difference between the measured energy loss in the TPC and the expected one of electrons at a given momentum is required to be within the mean energy loss measured for electrons and +3$\sigma$. This is shown for the most peripheral centrality class in Fig. \ref{fig:TPCpid}. The resolution on the energy loss in the TPC per unit length is between 5.5\% for peripheral and 6.35\% for central collisions. Since the contamination from misidentified hadrons is not negligible at low and high \pt, the analysis is performed in the \pt range \ptrange{1.5}{6}. In this \pt region the remaining background is estimated by a fit of multiple gaussian functions to the dE/dx-distribution in the TPC for given momentum intervals. The contamination is less than 10\% and it is subtracted. The \pt coverage will be increased both towards higher and lower \pt in the future. For this the Transition Radiation Detector (TRD) and the Electromagnetic Calorimeter (EMCal) will be included in the analysis. 

%Electron background
In addition to electrons from heavy-flavour hadron decays, the inclusive electron sample contains electrons from different background sources, i.e. photon conversions, Dalitz decays of light mesons (namely $\pi^{0}, \eta$ and $\eta^{\prime}$), dielectron decays of vector mesons ($\rho, \omega, \phi$), and virtual and real direct radiation. To account for this, a ``cocktail'' containing electrons from the various background sources is calculated. The cocktail is based on the measured charged pion spectrum assuming $\pi^0 = (\pi^+ +\pi^-)/2$. The spectrum is fitted with the sum of a modified Hagedorn function and an exponential, and this parameterization is used for the other mesons via \mt scaling. Not yet included are the heavy vector mesons $J/\psi$ and $\Upsilon$. The amount of electrons from conversions is calculated from the known material budget. For the contribution from direct radiation a pp NLO-calculation \cite{directphotons} is used, which is scaled to Pb--Pb applying binary collision scaling.

%Evaluation of the systematic errors
The systematic uncertainty is evaluated by varying track selection and particle identification cuts. To derive the total systematic error, the systematic errors of the single sources are added in quadrature. A total systematic error of 35\% independent on momentum was obtained. The largest contribution to the systematic error is the electron selection in the TPC.

The systematic error of the electron cocktail was calculated by changing the cocktail inputs by their respective uncertainties. As done for the systematic uncertainty of the inclusive electron spectrum, the systematic uncertainties of the individual sources are added in quadrature to derive the total systematic uncertainty, which is $\approx 25\%$. The main contribution to the systematic uncertainty arises from the systematic uncertainty on the charged pion differential cross section used to evaluate the Dalitz and conversion electron distribution.

%Reference measurement in proton-proton collisions
The nuclear modification factor is defined as
\begin{equation}
R_{AA} = \frac{1}{<T_{AA}>}\frac{\mathrm{d}N_{AA}/\mathrm{d}p_{\mathrm t}}{\mathrm{d}\sigma_{pp}/\mathrm{d}p_{\mathrm t}}
\end{equation}
where $<T_{AA}>$ is the average of the nuclear overlap function and $d\sigma/dp_{\mathrm t}$ is the differential cross section in proton-proton collisions. Electrons from heavy-flavour decays are measured in ALICE in proton-proton collisions at \roots = 7 \TeV \cite{sma_qm}. The analysis is done in the same way as for Pb--Pb-collisions, however the Transition Radiation Detector (TRD) is used in addition to the TOF and TPC for the electron selection. This allows a measurement up to a \pt of 10 \GeVc. 

\begin{figure}[t]
\includegraphics[width=0.7\textwidth]{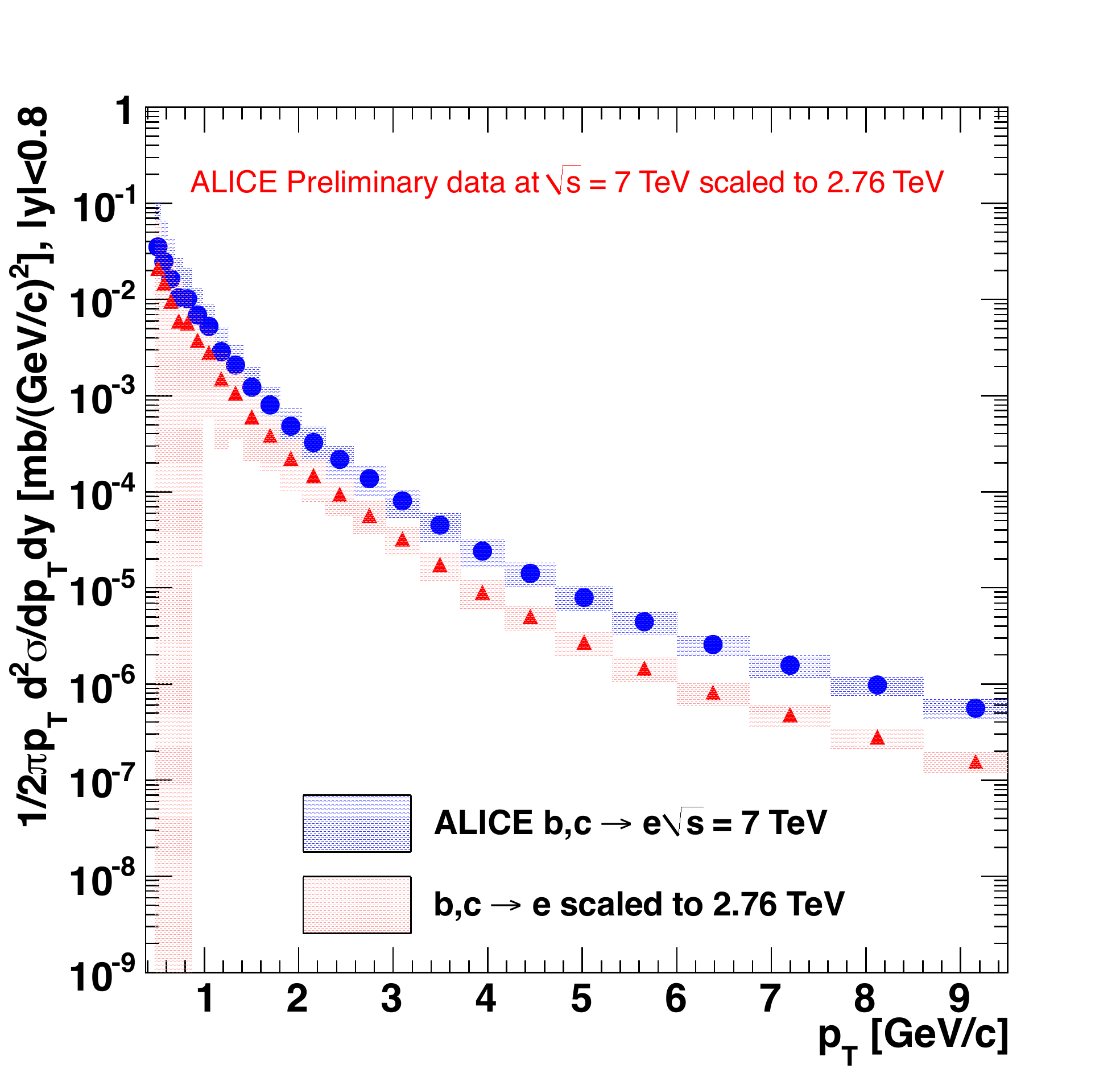}
\caption{Heavy-flavour electron \pt spectrum measured at \roots = 7 \TeV (blue) and the spectrum for \roots = 2.76 \TeV, obtained by scaling the one at \roots = 7 \TeV, based on FONLL calculations. Boxes indicate the systematic uncertainty.}
\label{fig:ppref}
\end{figure}
The reference \pt-spectrum at \roots = 2.76 \TeV, shown in Fig. \ref{fig:ppref}, is derived from the measured heavy-flavour electron spectrum at \roots = 7 \TeV by a scaling based on Fixed Order plus Next-to-Leading-Logarithms (FONLL) pQCD calculations\cite{Cacciari:1998it} (scaling method described in \cite{Averbeck:2011ga}). The additional uncertainty arising from the scaling procedure is small.

\section{Results}
\label{sec:res}
% Inclusive over cocktail
\begin{figure}[t]
\centering
\includegraphics[width=0.7\textwidth]{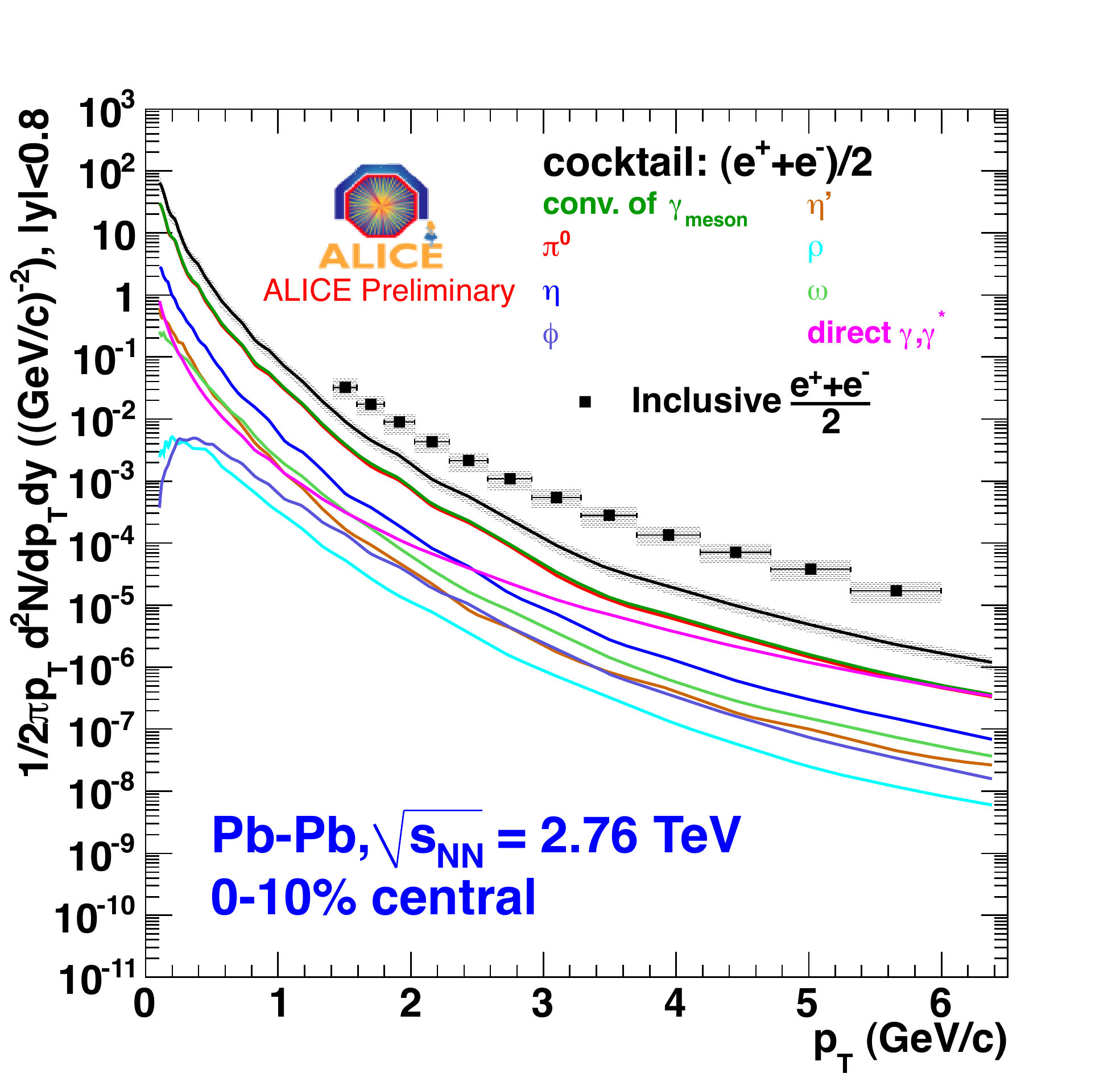}
\caption{Inclusive electron spectrum for the most central Pb--Pb events (0 - 10 \%) compared to the cocktail of electrons from background sources. An excess over cocktail is visible.}
\label{fig:incEl}
\end{figure}

\begin{figure}[t]
\includegraphics[width=0.35\textwidth]{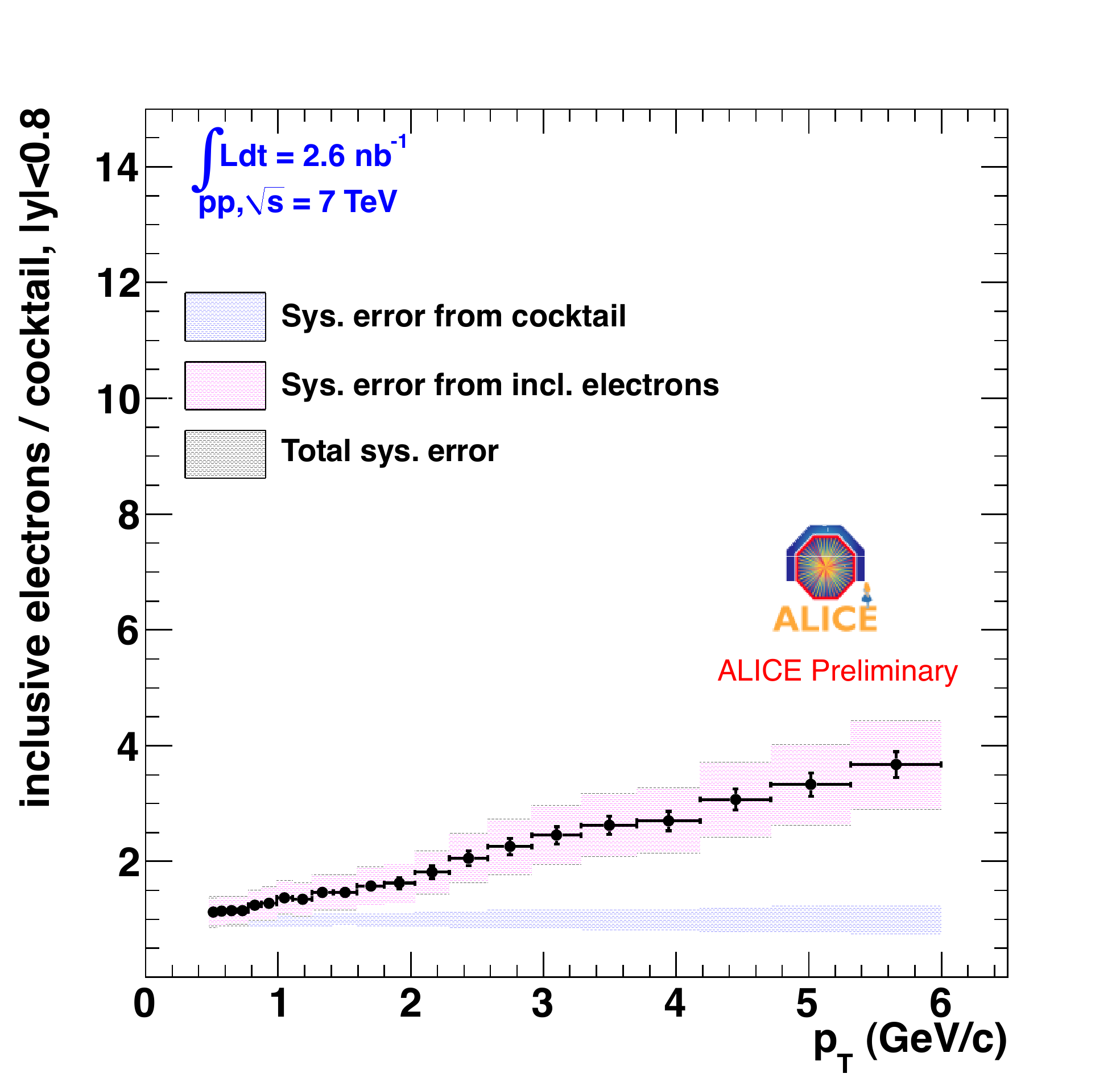}
\includegraphics[width=0.35\textwidth]{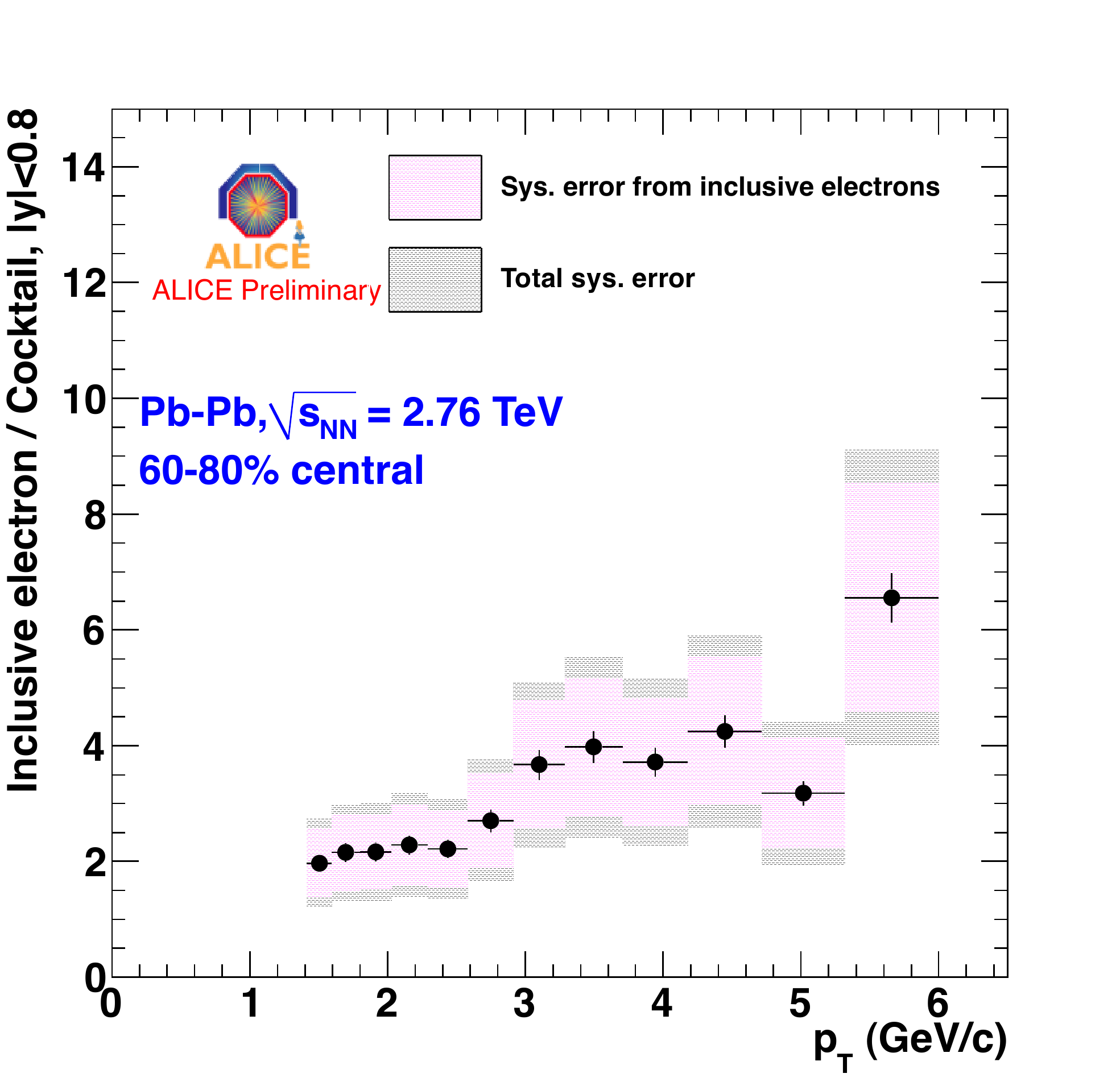}
\includegraphics[width=0.35\textwidth]{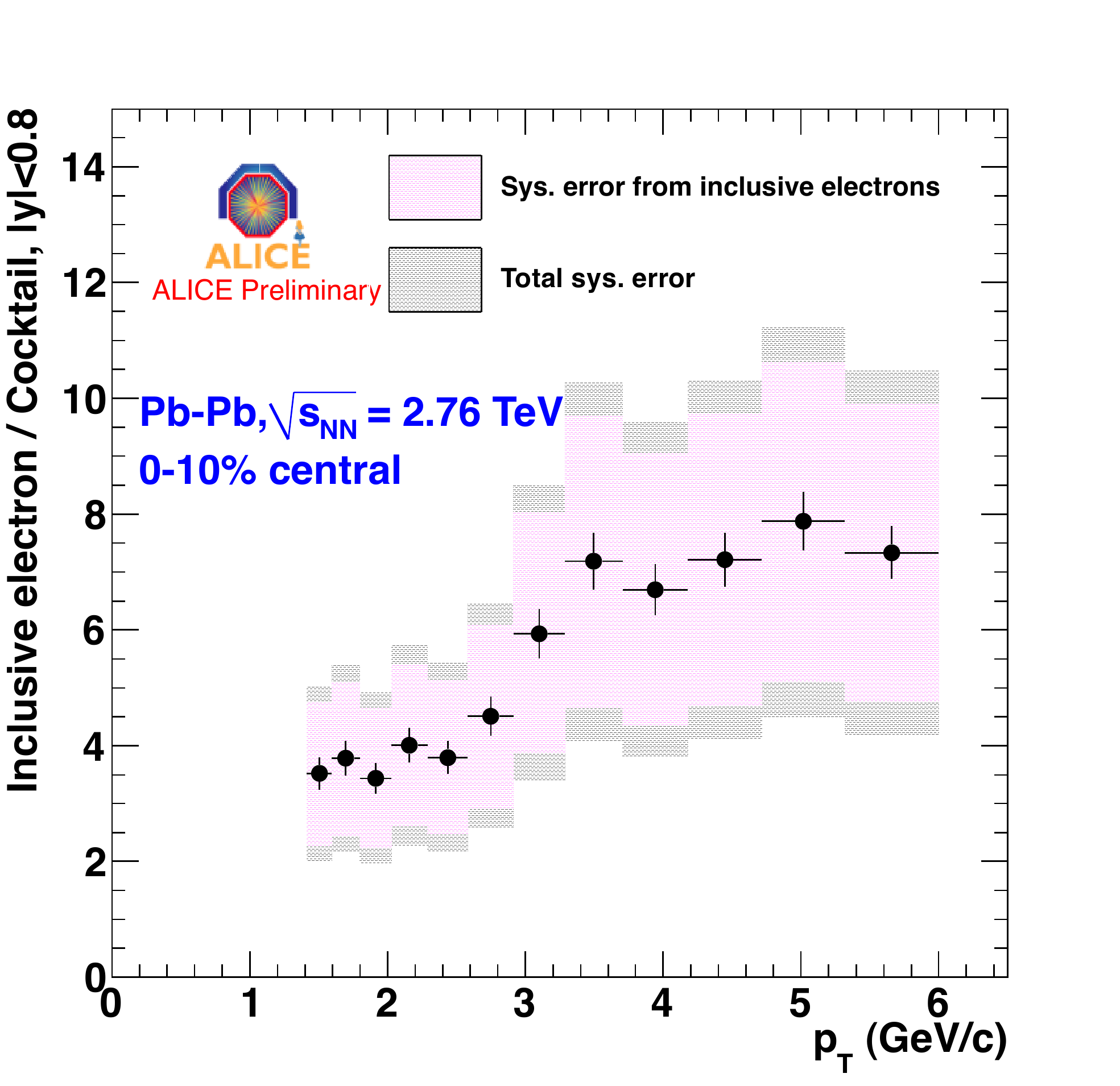}
\caption{Ratio of inclusive electrons over cocktail in proton-proton collisions (left), peripheral Pb--Pb (middle) and central Pb--Pb collisions (right).}
\label{fig:incCock}
\end{figure}

The inclusive electron spectrum compared to the cocktail is shown for the most central Pb--Pb collisions(0-10\%) in Fig. \ref{fig:incEl}. An excess over the cocktail is visible. This excess is interpreted in terms of semi-electronic heavy-flavour decays for \pt > 3.5 \GeVc. In addition, there is a hint for an excess from a different source in the \pt region below about 3.5 \GeVc. In Fig. \ref{fig:incCock} the ratio of inclusive electrons to the cocktail is shown for proton-proton collisions, peripheral(60-80\%), and central (0-10\%) Pb--Pb collisions. The inclusive electron spectrum in proton-proton collisions is dominated by electrons from background sources towards low \pt. Therefore, this ratio approaches unity towards zero. In Pb--Pb collisions this ratio is larger than observed in proton-proton collisions, especially in the most central collisions. A possible explanation for the excess relative to the proton-proton case could be thermal radiation. In the \pt region \ptrange{3.5}{6} the difference between inclusive electrons and the cocktail is expected to be dominated by electrons from heavy-flavour hadron decays. 

% RAA
\begin{figure}
\begin{minipage}{0.55\textwidth}
\includegraphics[width=\textwidth]{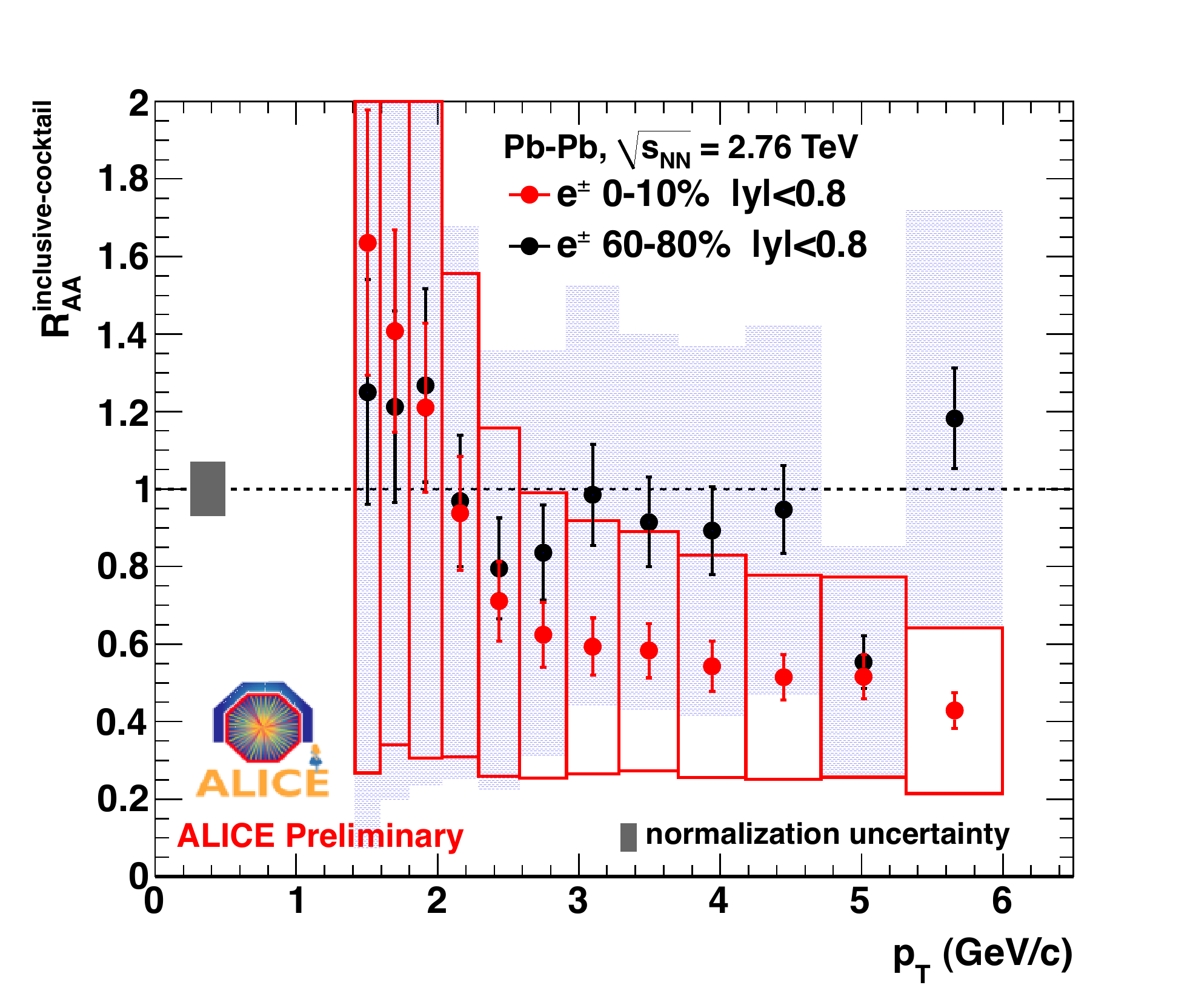}
\end{minipage}
%\hspace{0.02\textwidth}
\begin{minipage}{0.55\textwidth}
\includegraphics[width=\textwidth]{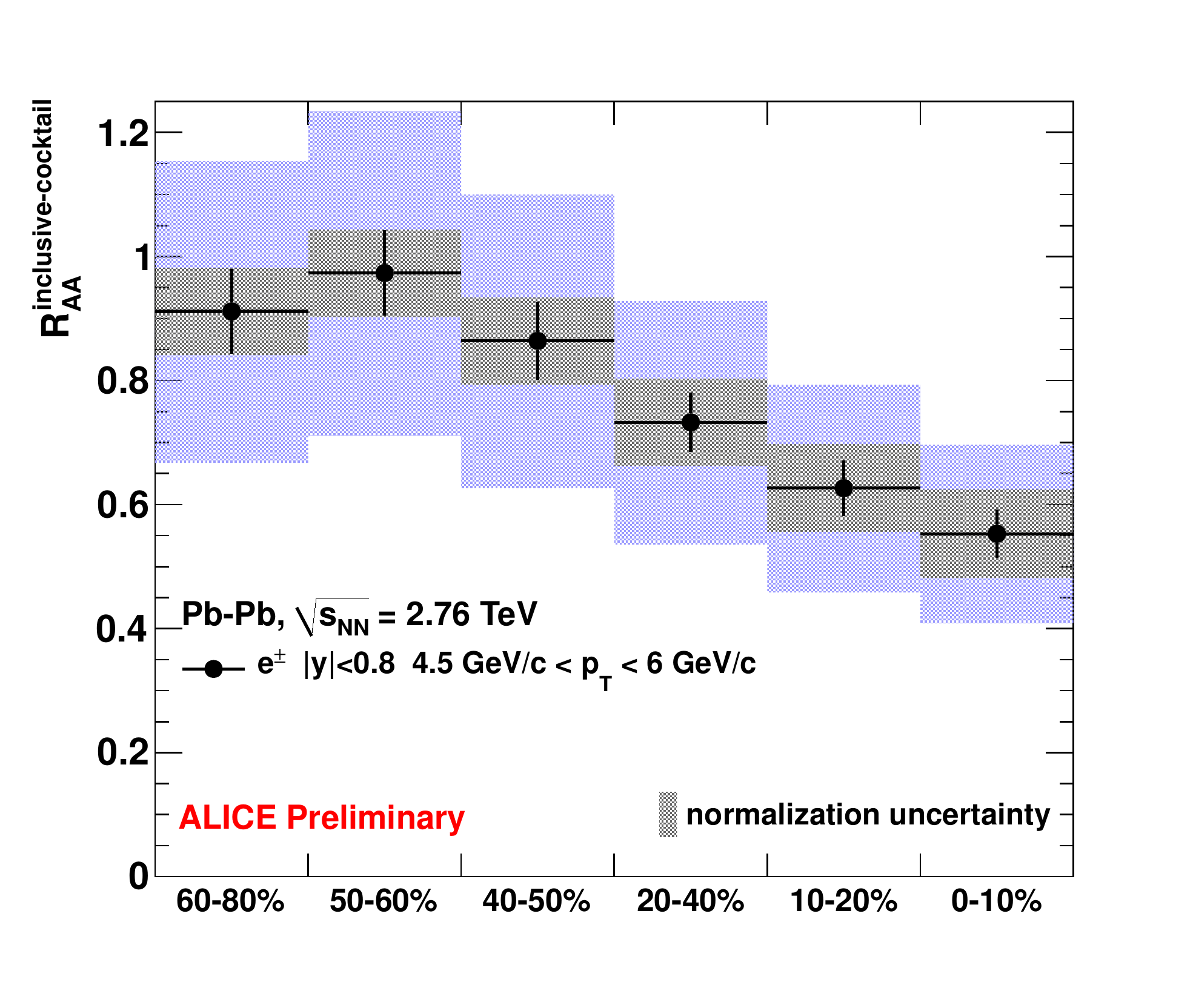}
\end{minipage}
\caption{Nuclear modification factor of cocktail-subtracted  electrons: on the left side the nuclear modification factor is shown for the most peripheral 60-80\% (black points) and the most central 0-10\% Pb--Pb collisions (red points). Boxes indicate the systematic uncertainty. The dependence of the nuclear modification factor on collision centrality for \ptrange{4.2}{6} is shown on the right side.}
\label{fig:raaHFE}
\end{figure}
The nuclear modification factor of cocktail-subtracted  electrons is shown in Fig. \ref{fig:raaHFE}. On the left side the nuclear modification factor is shown for two centrality classes, the most peripheral one (60-80\%) and the most central one (0-10\%). A suppression is seen in the high-\pt region for the most central collisions, although systematic uncertainties are still large at this moment. On the right side of Fig. \ref{fig:raaHFE} the nuclear modification factor is shown in the \pt region \ptrange{4.2}{6} as a function of centrality, indicating that the suppression increases with centrality.

\section{Conclusions and outlook}
We presented the measurement of the nuclear modification factor of electrons from heavy-flavour hadron decays in Pb--Pb collisions. A suppression is seen for the most central collisions. In the \pt region \ptrange{3.5}{6} the suppression increases with centrality. In the \pt region \ptrange{1.5}{3.5} a hint for an excess over the cocktail is seen, which is not visible in proton-proton collisions and which is increasing with centrality. A possible source for this excess could be a contribution from thermal radiation.

\section{Acknowledgements}  
Copyright CERN for the benefit of the ALICE Collaboration. 
%%%%%%%%%%%%%%%%%%%%%%%%%%%%%%%%%%%%%%%%%%%%%%%%
%% BACKMATTER
%%%%%%%%%%%%%%%%%%%%%%%%%%%%%%%%%%%%%%%%%%%%%%%%

%\begin{theacknowledgments}
%\end{theacknowledgments}



\bibliographystyle{aipproc}   % if natbib is available

\bibliography{proceedings_hfe}

\IfFileExists{\jobname.bbl}{}
 {\typeout{}
  \typeout{******************************************}
  \typeout{** Please run "bibtex \jobname" to optain}
  \typeout{** the bibliography and then re-run LaTeX}
  \typeout{** twice to fix the references!}
  \typeout{******************************************}
  \typeout{}
 }

\end{document}